\documentclass[jcp,aip,
 amsmath,amssymb,
reprint,
]{revtex4-1}

\usepackage{graphicx,subcaption}% Include figure files
\usepackage{dcolumn}% Align table columns on decimal point
\usepackage{comment}
\usepackage{bm}% bold math
\usepackage[utf8]{inputenc}
\usepackage[T1]{fontenc}
\usepackage{mathptmx}
\usepackage{etoolbox}
\usepackage{mathrsfs}
\usepackage{xcolor}

\makeatletter
\def\@email#1#2{%
 \endgroup
 \patchcmd{\titleblock@produce}
  {\frontmatter@RRAPformat}
  {\frontmatter@RRAPformat{\produce@RRAP{*#1\href{mailto:#2}{#2}}}\frontmatter@RRAPformat}
  {}{}
}%
\makeatother
\begin{document}

\title{Free-energy REconstruction from Stable Clusters (FRESC): A new method to evaluate nucleation barriers from simulation}

\author{Adrián Llamas-Jaramillo}
\affiliation{ 
Departament de Física de la Matèria Condensada, Universitat de Barcelona, 08028 Barcelona, Spain.\\}
 
\author{Ivan Latella}%
\affiliation{ 
Departament de Física de la Matèria Condensada, Universitat de Barcelona, 08028 Barcelona, Spain.\\}
\affiliation{Universitat de Barcelona, Institute of Nanoscience and Nanotechnology (IN2UB), 08028 Barcelona, Spain.\\}

\author{David Reguera}
\email{dreguera@ub.edu.}

\affiliation{ 
Departament de Física de la Matèria Condensada, Universitat de Barcelona, 08028 Barcelona, Spain.\\}
\affiliation{Universitat de Barcelona, Institute of Complex Systems (UBICS), 08028 Barcelona, Spain.\\}

\date{\today}

\begin{abstract}
We present a 
simulation technique to evaluate the most important quantity for nucleation processes: the nucleation barrier, i.e. the free energy of formation of the critical cluster. The method is based on stabilizing a small cluster by simulating it in the $NVT$ ensemble and using  the thermodynamics of small systems to convert the properties of this {\it stable} cluster into the Gibbs free energy of formation of the {\it critical} cluster. 
We demonstrate this approach using condensation in a Lennard-Jones truncated and shifted fluid as an example, showing an excellent agreement with previous Umbrella Sampling simulations. 
The method is straightforward to implement, computationally inexpensive, requires only a small number of particles comparable to the critical cluster size, does not rely on the use of Classical Nucleation Theory, and does not require any cluster definition or reaction coordinate. 
All of these advantages hold the promise of opening the door to simulate nucleation processes in complex molecules of atmospheric, chemical or pharmaceutical interest that cannot be easily simulated with current  
techniques.

\end{abstract}

\maketitle

\section{\label{sec:intro} Introduction\\}

Nucleation is the process controlling first order phase transitions and it is crucial for a wide variety of phenomena of the utmost current interest. Examples range from cloud formation and crystallization in materials science to protein aggregation in biological systems. The common feature of all nucleation phenomena is the existence of an energy barrier that must be overcome in order to trigger the formation of the new phase. This barrier arises from the competition between the energetic cost of creating an interface and the thermodynamic driving force favoring the new phase. Understanding the microscopic mechanisms that govern nucleation is essential for predicting and controlling phase transitions in both natural and engineered systems.

For the sake of simplicity, we will focus on the simplest example of a nucleation phenomenon: the homogeneous condensation of a single component vapor. In this case, the qualitative physical picture is well known \cite{Abraham1974,Laaksonen1995,Debenedetti1996,Kashchiev2000,Vehkamki2006}. When a vapor has a pressure above the equilibrium saturation pressure, it becomes supersaturated and the thermodynamic stable state is the liquid phase. The transformation of the supersaturated vapor into the liquid starts with the formation, by thermal fluctuations, of small aggregates of molecules 
having liquid-like properties.  These clusters enjoy the energetic advantage of having a smaller chemical potential (i.e. energy per molecule of the more stable liquid phase), but must pay the price of the surface energy associated to the presence of an interface between the incipient liquid and the vapor. The combination of these two opposite contributions gives rise to an energy barrier, whose maximum is located at a particular cluster size, known as the critical cluster size.  
In the context of Classical Nucleation Theory (CNT) \cite{Becker1935a,zeldovich1942contribution,farkas1927keimbildungsgeschwindigkeit,Volmer1926a,Abraham1974,Debenedetti1996,Kashchiev2000,Vehkamki2006,Laaksonen1995}, the work of formation of this critical cluster, $\Delta G^*$ or $\Delta \Omega^*$, is simply given by\cite{Gibbs} $\Delta G^*= \Delta \Omega^*= 16 \pi \gamma^3/3 (\Delta p)^2$, where $\Delta p= p_l (\mu, T) - p_v (\mu, T) $ is just the pressure difference between the liquid and the vapor phase at the same chemical potential $\mu$ and temperature $T$, and $\gamma$ is the surface tension of the liquid drop defined at the surface of tension. The radius of the critical cluster is $R^*= 2 \gamma/\Delta p$, and  the nucleation rate at which the phase transformation is initiated is given by $J= K \exp (-\Delta \Omega^*/k_B T)$, where $K$ is a kinetic prefactor and $k_B$ is Boltzmann's constant. This exponential dependence of the nucleation rate on the free energy barrier is the distinctive trait of activated processes and makes the prediction of nucleation phenomena specially challenging. Any small inaccuracy in the evaluation of $\Delta \Omega^*$ leads to orders of magnitude discrepancies between the predictions and experiments or simulations\cite{iland2007argon,Wedekind2007c}. 

Accurately evaluating the nucleation barrier or nucleation rate remains a challenging task due to the intrinsic features of nucleation phenomena. The formation of a critically sized cluster is stochastic and becomes increasingly rare as the barrier height grows. Even when such a cluster does appear, it is inherently unstable because its size corresponds to the top of the free-energy barrier. Consequently, it does not persist long enough to reliably measure its properties.

Despite these difficulties, numerous simulation approaches
have been developed to study nucleation processes. 
The most straightforward are direct or brute force simulations\cite{Wedekind2007c,Diemand2014}, 
where a supersaturated system is prepared and the spontaneous emergence of large clusters is monitored. Nucleation rates or cluster free energies can then be estimated
using different analysis 
schemes
like the Mean First Passage Time \cite{Wedekind2007a,Wedekind2008a}. However, since the appearance of a critical cluster is a rare event, this 
method
requires simulating at very high supersaturations, far from those experimentally accessible (except for Laval nozzle experiments \cite{Wyslouzil2016,Li2021}).  
To overcome this difficulty, a variety of rare-event simulation approaches have been proposed, including Transition Path Sampling~\cite{Dellago2003,Bolhuis2002}, Forward Flux Sampling~\cite{Allen2009} and related variants. Nevertheless, these methodologies require the generation of a large ensemble of trajectories and a proper choice of a reaction coordinate that makes them  computationally demanding. 
Other types of techniques aim at obtaining the free energy landscape by adding a potential iteratively, like in Metadynamics\cite{Barducci2011} or a constraining parabolic potential, like in Umbrella Sampling \cite{Torrie1977,Frenkel2002}, in order to sample efficiently the whole free energy landscape or stabilize locally clusters of different sizes and reconstruct the free energy landscape. 
A recent alternative is the seeding technique \cite{Espinosa2016a,Rosales-Pelaez2019}, based on CNT, where one inserts a pre-formed cluster or seed in the simulation, monitors its time evolution, and tries to identify the critical cluster size as the seed having a $50 \%$ chance of growing or disappearing. 
Although all these methods have been successfully applied to analyze different nucleation phenomena, their implementation remains nontrivial and often computationally intensive due to the rare and unstable nature of the critical clusters, whose properties are the cornerstone of nucleation.

\subsection{Stable clusters in the canonical ensemble}

\begin{figure}[h!]
\centering
\includegraphics[width=\columnwidth]{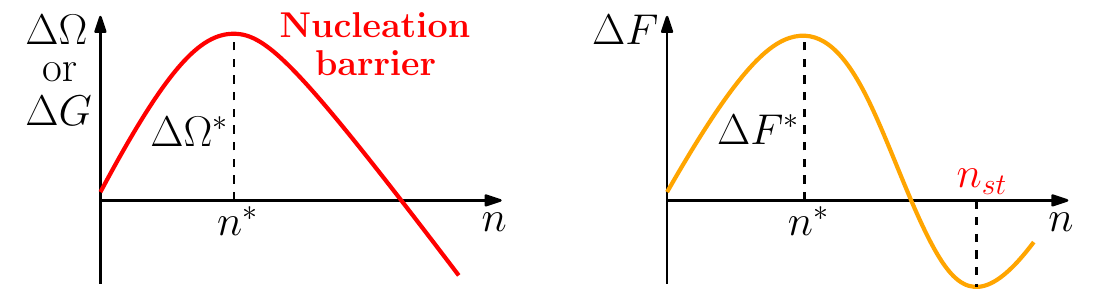}
\caption{\footnotesize{Comparison between the free energy landscape of formation of a cluster of size $n$ in the $\mu VT$ or $NPT$ ensembles, and the canonical $NVT$ ensemble. In the $\mu VT$ or $NPT$ ensembles there is only one extremum, a maximum in the associated free energy $\Delta \Omega$ or $\Delta G$, corresponding to the critical nucleus. Contrarily, in the $NVT$ ensemble at similar conditions, there may be two extrema, a maximum and a minimum in $\Delta F$, corresponding to the critical cluster and a stable cluster, respectively.}}
 \label{fig:landscape}
\end{figure}

A practical strategy to address
these difficulties is to stabilize the clusters of interest for nucleation. There is indeed a way to generate a small liquid cluster that is perfectly stable coexisting with its vapor, by working in the canonical $NVT$ ensemble. When nucleation is studied experimentally or in simulations in the $\mu VT$ or $NPT$ ensemble, the critical cluster is unstable and the free energy landscape for cluster formation only has a maximum, as shown schematically in Fig.~\ref{fig:landscape}. However, the same system in the canonical ensemble has a maximum, corresponding to the critical size for the formation of this cluster in the canonical ensemble, but also can have a local minimum corresponding to a stable (or metastable) cluster.

The fact that a small liquid-like cluster can be stabilized in the canonical ensemble is not new. Since the seminal simulations by Lee, Barker, and Abraham\cite{Lee1973}, Binder \cite{Binder1975}, and Rao, Berne and Kalos \cite{Rao1978}, or the theoretical works by Yang \cite{Yang1983,Yang1985}, it is known that clusters can be stabilized in the canonical ensemble. Density Functional Theory studies by Lee, Telo da Gamma, and Gubbins\cite{Lee1986},  and Talanquer and Oxtoby\cite{Talanquer1994} also  showed the stability of clusters in the canonical ensemble. 

A simple way to quantify and analyze the stability of small liquid clusters in the canonical ensemble is to use the Modified Liquid Drop model\cite{Weakliem1993,Reguera2003}.  
In CNT, the critical cluster is defined by the size at which the cluster is
in chemical and mechanical equilibrium with the vapor for a given set of thermodynamic conditions. Therefore, it fulfills the requirements of equal chemical potentials
\begin{equation}
    \mu_l(T,p_l) =\mu_v(T,p_v)
\label{Eq:equal_chem}    
\end{equation}
\noindent and the Laplace relation
\begin{equation}
     p_l - p_v = \frac{2\gamma}{R} .
\label{Eq:Laplace}    
\end{equation}
 In systems where either the pressure or the chemical potential are the control variables, there is only one 
 non-uniform solution to these two equations, corresponding to the critical cluster, which is the maximum of the corresponding Gibbs free energy $\Delta G(n)$ or grand potential $\Delta \Omega(n)$ work of formation as a function of the cluster size $n$. However, in the canonical $NVT$ ensemble, under appropriate conditions, there turn out to be two solutions fulfilling these conditions: one corresponding to the critical cluster and a second one corresponding to a {\it stable} or metastable cluster. Both are extrema of the Helmholtz free energy landscape $\Delta F(n)$, but the critical cluster is a maximum, whereas the stable cluster corresponds to a (local) minimum (see Fig.~\ref{fig:landscape}).
 
The 
reason why a cluster can be stabilized in the $NVT$ ensemble is 
straightforward.
Since the total number of molecules $N$ and the volume $V$ are  fixed, in a supersaturated vapor the liquid cluster grows at the expense of the vapor molecules, reducing the supersaturation and eventually reaching a situation where the cluster cannot grow further. 
What is important is that the resulting stable cluster also fulfills Eqs.~(\ref{Eq:equal_chem}) and (\ref{Eq:Laplace}).
 However, it
is surrounded by a vapor at a vapor pressure $p_v$ and chemical potential $\mu$ that are not dictated by the homogeneous density $N/V$, but are rather determined by the remaining vapor molecules. 
If this cluster in equilibrium with the vapor is placed in a system either at a fixed chemical potential $\mu$, corresponding to that of the vapor instead of a constant $N$, or at fixed vapor pressure $p$ instead of a fixed $V$, then it would correspond to the critical cluster since it is the only solution to the equilibrium conditions Eqs.~(\ref{Eq:equal_chem}) and (\ref{Eq:Laplace}) in the $\mu VT$ or $NPT$ ensemble. The fact that the stable cluster in the canonical ensemble is equivalent to the critical cluster in the $\mu VT$ or $NPT$ ensembles has been recognized several times in the literature~\cite{Talanquer1994,Reguera2003} and verified in simulations~\cite{Rosales-Pelaez2020,MonteroDeHijes2020,Sanchez-Burgos2020,MonteroDeHijes2020a}.

Therefore a promising strategy for evaluating the properties of the critical cluster is 
to generate
a stable cluster in the $NVT$ ensemble and measure its properties, 
particularly
its free energy of formation. 
Numerous studies in the literature have pursued this strategy by simulating stable clusters in the canonical ensemble.
A seminal 
contribution was provided
by Lee, Barker and Abraham\cite{Lee1973} who performed Monte Carlo simulations with less than 100 Lennard-Jones molecules at different volumes. They evaluated the Helmoltz free energy using thermodynamic integration, but treating the whole $NVT$ system as a physical cluster. MCGuinty\cite{McGinty1973}, Rao, Berne and Kalos\cite{Rao1978}, and Rusanov and Brodskaya\cite{Rusanov1977} also simulated and analyzed the properties of stable liquid-like clusters, but not directly its free-energy of formation. Binder and coworkers \cite{Binder1975,Binder2003,MacDowell2004,MacDowell2006,Schrader2009,Block2010,Binder2012,troster2012numerical,Statt2015,Troster2018,Ko2018} developed a 
comprehensive series of studies addressing
the stability of spherical and cylindrical drops and bubbles in the canonical ensemble. 
They introduced several strategies to evaluate the properties of stable clusters, including their free energy of formation and surface tension, using combinations of finite-size scaling, successive umbrella sampling, and histogram reweighting techniques. In particular, many of these approaches
rely on identifying the size of the stable cluster using a suitable reaction coordinate and are computationally demanding or not easy to implement. Recently, the seeding technique developed by C. Vega and coworkers has also been extended to the $NVT$ ensemble \cite{Rosales-Pelaez2020,MonteroDeHijes2020}. Another remarkable simulation techniques are the gauge cell and ghost field methods introduced by Neimark and collaborators~\cite{Neimark2000,Neimark2005c,Neimark2005e,Neimark2006b}.
To evaluate the nucleation barrier, they consider the addition of
a second reservoir whose size controls the strength of fluctuations and the stability of the system, and an external ("ghost") field to control the formation of a droplet with the desired size and density. Finally, very recently, 
the Modified Liquid Drop Model has been combined with the correspondence between a stabilized cluster in the canonical ensemble and the critical cluster relevant for nucleation, enabling the evaluation of nucleation barriers and rates under the assumption of CNT validity~\cite{Li2023,Li2025}.

Building on the previous insightful works, in this article, we present a different simulation technique that we call FRESC: Free-energy REconstruction from Stable Clusters. This technique is 
straightforward
to implement, since it only requires a set of standard $NVT$ Monte Carlo (MC) or Molecular Dynamics (MD) simulations; it is very efficient, as
it only involves a small number of molecules of the same order as the size of the critical cluster; and it allows for
an accurate evaluation of the nucleation barrier, i.e. the free energy of formation of the critical cluster, at any value of the supersaturation. In particular,
evaluating the nucleation barrier near the spinodal is notoriously difficult using standard techniques such as umbrella sampling;
with FRESC it can be obtained using simulations with
few molecules.

We illustrate this method for the particular case of condensation using a simple Lennard-Jones Truncated and Shifted potential. But it can be applied to any intermolecular potential, and extended to other nucleation phenomena 
such as
cavitation, bubble formation, or crystallization in single or multicomponent systems.

\section{\label{sec:method}FRESC Method}

The idea of the method is extremely simple. First, a standard MC or MD simulation of a supersaturated system is performed in the canonical $NVT$ ensemble. From this simulation, the Helmholtz 
free energy of formation of the  
stable cluster can be obtained via straightforward thermodynamic integration in the way described below.

In the canonical ensemble, the Helmholtz free energy of the whole system $F(N,V,T)$ satisfies
\begin{equation}
    dF = - S dT - p dV +\mu dN,
    \label{eq:dF}
\end{equation}
where $S$, $p$, and $\mu$ are the corresponding entropy, pressure, and chemical potential, respectively.
Hence, at constant temperature $T$, there are two possible ways to obtain the Helmholtz free energy: either we fix the total number of molecules $N$, perform a set of simulations at different volumes $V'$, measure the pressure at each simulation $p(N,V',T)$, and get 
\begin{equation}
    F(N,V,T) = - \int_{V_{max}}^V p(N,V',T) dV' + F(N,V_{max},T)  
    \label{eq:F_V}
\end{equation}
or we fix the volume $V$, perform a set of simulations using different numbers of molecules $N'$ measuring the chemical potential $\mu(N',V,T)$, and get
\begin{equation}
    F(N,V,T) = \int_{N_{min}}^N \mu(N',V,T) dN' + F(N_{min},V,T).  
    \label{eq:F_N}
\end{equation}
Here, $F(N,V_{max},T)$ and $F(N_{min},V,T)$ are the Helmholtz free energies of the system at a reference volume $V_{max}$ and a reference number of particles $N_{min}$, respectively. 
The Helmholtz free energy of the stable cluster is simply the difference between the free energy of the system containing the cluster, i.e. $F(N,V,T)$, and that of the uniform vapor $F_0(N,V,T)$ with density $\rho_0=N/V$, namely, 

\begin{equation}
     \Delta F(N,V,T) = F(N,V,T) - F_0(N,V,T).
    \label{eq:DeltaF}  
\end{equation}

The grand potential $\Delta \Omega^*$ or Gibbs free energy $\Delta G^*$ of the critical cluster can be obtained from $\Delta F$ by exploiting the thermodynamic transformation\cite{Lee1986,Reguera2004,Reguera2004c,Wilhelmsen2015c}
\begin{equation}
    \Delta \Omega^*(N,V,T) = \Delta F(N,V,T) + V \Delta p -N \Delta \mu  ,  
    \label{eq:DeltaG_definition}
\end{equation}
where $\Delta p = p -p_0$ and $\Delta \mu = \mu -\mu_0$, with $p_0$ being the pressure of the uniform vapor with density $\rho_0=N/V$ and chemical potential $\mu_0$.
Equation~(\ref{eq:DeltaG_definition}) was rigorously derived in the appendix of Ref.~\cite{Reguera2004c} and can also be obtained using the thermodynamics of small systems~\cite{Hill}. The values of $p$ and $\mu$ in this equation are obtained from the MC or MD simulation in the $NVT$ ensemble with $N$ particles, whereas $p_0$ and $\mu_0$ are the properties of the reference system in the uniform vapor phase.

In principle, it seems that an accurate equation of state (EoS) would be required to account for $p_0$ and $\mu_0$ in the uniform supersaturated vapor phase. But there is a way to avoid the use of the EoS. We first note that in a system with a small number of particles, there might be nonextensive corrections even for the ideal gas~\cite{Hill}. A way to minimize these corrections is working with excess quantities, i.e. by subtracting from free energies, pressures, and chemical potentials the corresponding ideal contributions. Denoting by $F^{ex}$, $p^{ex}$, and $\mu^{ex}$ the excess free energy, pressure, and chemical potential of the system with the cluster, respectively, and by $F_0^{ex}$, $p_0^{ex}$, and $\mu_0^{ex}$ those for the uniform vapor with density $\rho_0=N/V$, from Eq.~(\ref{eq:DeltaG_definition}) we get
\begin{eqnarray}
    \Delta \Omega^*(N,V,T) &=& F^{ex}(N,V,T) + V p^{ex} -N \mu^{ex}  \nonumber\\
    &-& \left[ F_0^{ex}(N,V,T) + V p_0^{ex} -N \mu_0^{ex}\right].
    \label{eq:DeltaG_definition_2}
\end{eqnarray}
Assuming now that the standard thermodynamic relation $F_0^{ex} =-p_0^{ex}V +\mu_0^{ex} N$ applies for the homogeneous vapor, Eq.~(\ref{eq:DeltaG_definition_2}) simply becomes
\begin{equation}
    \Delta \Omega^*(N,V,T) = F^{ex}(N,V,T) + p^{ex} V - \mu^{ex} N,    
    \label{eq:DeltaG}
\end{equation}
with the advantage that no EoS or previous evaluation of the properties of the homogeneous metastable phase is needed. 
Notice that the relation $F_0^{ex} =-p_0^{ex}V +\mu_0^{ex} N$ is an approximation strictly valid only in the 
thermodynamic limit. For very small systems, 
even if a cluster is not formed, an inhomogeneity can also arise in the vapor phase,  
for instance, due to boundary effects.

Out of the two strategies to get the Helmholtz free energy $F(N,V,T)$ by thermodynamic integration, it is more convenient to integrate the chemical potential, Eq.~(\ref{eq:F_N}). One advantage of this procedure is that, unlike the local pressure, the chemical potential $\mu$ is uniform throughout the system even in the presence of the stable drop. Moreover, in the  vapor phase surrounding the liquid drop, it can be measured particularly easily using, for example, the Widom insertion method~\cite{Widom1963,Widom1978,Frenkel2002,Perego2018}. A second advantage is computational: since a set of simulations is required, using very small number of particles in the simulations to compute the integral at fixed volume is more efficient than doing all simulations with the same $N$ at different volumes.

Taking into account these considerations, by combining Eq.~(\ref{eq:F_N}) expressed in terms of excess quantities and Eq.~(\ref{eq:DeltaG}), the final expression to evaluate the free energy of formation of critical clusters in the FRESC method is just
\begin{eqnarray}
     \Delta \Omega^*(N,V,T) =&&  \int_{N_{min}}^N \mu^{ex}(N') dN'  + p^{ex}(N) V - \mu^{ex}(N) N \nonumber\\&&
     +F^{ex}(N_{min},V,T),
    \label{eq:DeltaG_simu} 
\end{eqnarray}
where $N_{min}$ corresponds to the smallest system size to simulate, that must correspond to a homogeneous vapor state for which $F^{ex}(N_{min},V,T) = - p^{ex}(N_{min}) V + \mu^{ex}(N_{min}) N_{min}$. 
In practice, choosing a very small value of $N_{min}$ would yield a value of $F^{ex}(N_{min},V,T)$ 
essentially
zero. 

It is worth stressing that the FRESC method does not rely on the use or validity of CNT, and no cluster criteria, reaction coordinate or explicit cluster identification is required. The critical cluster size can be evaluated directly from the excess, as explained in the Appendix. If, at the chosen conditions, no cluster is formed, then the free energy of formation Eq.~(\ref{eq:DeltaG_definition_2}) is simply zero, as it would correspond to a homogeneous system. Notice also that, when applied to systems with different $N$,  this equation will provide the nucleation barrier for a range of different critical cluster sizes and conditions.

\section{\label{sec:MC} Monte Carlo simulations}

In order to demonstrate this novel technique, 
we implemented a standard~\cite{Frenkel2002} 
MC
simulation in the canonical ensemble for a system of particles interacting with a Lennard-Jones 
(LJ)
potential truncated and shifted at a reduced cut-off radius of $r_c=2.5\sigma$, LJTS($2.5\sigma$), 
expressed as
\begin{equation}
    \begin{split}
     U_{\text{LJTS}}(r_{ij}) &=
    \begin{cases}
        U_{\text{LJ}}(r_{ij})-U_{\text{LJ}}(r_c), & \text{if} \ r_{ij} \leq r_c\\
        0, & \text{if} \ r_{ij} > r_c
    \end{cases}
    \end{split}.
\label{eq:LJTS}
\end{equation}
Here,
\begin{equation}
    U_{\text{LJ}}(r_{ij}) = 4\varepsilon\bigg[\bigg(\frac{\sigma}{r_{ij}}\bigg)^{12}-\bigg(\frac{\sigma}{r_{ij}}\bigg)^6\bigg]
\label{eq:LJ}
\end{equation}
\noindent is the standard 
LJ
potential with $r_{ij}$ being the distance between two particles, $\sigma$  the distance at which the potential is zero, and $\varepsilon$ the energy minimum of the potential well.

Simulations were performed in reduced units in terms of the LJ particle's mass $m$, distance $\sigma$, and energy $\varepsilon$, at a reduced temperature $T=0.625$.
The simulation box is spherical, without periodic boundary conditions and with a fixed center of mass. For each simulation, the initial configuration 
is chosen as
an fcc lattice and we correct its center of mass such that it coincides with the desired origin of coordinates (center of the spherical simulation box). 

Once the system is initialized, for each MC step, a randomly selected particle is displaced a random distance and another randomly chosen particle is displaced by the \textit{same} amount but in opposite direction, which maintain the center of mass fixed. This trial displacement of the two particles is accepted or rejected according to the standard Metropolis algorithm. The maximum displacement is adjusted during equilibration to get an average acceptance of  $50\%$ of the trial moves. In each simulation, $10^7$ equilibration and $10^7$ production steps were performed to get 
reliable
statistics. 

The pressure $p$ is calculated by averaging the virial. In order to measure the chemical potential $\mu$, we implement the Widom insertion method for a spatially inhomogeneous system~\cite{Widom1978,Perego2018} in which the excess chemical potential is calculated as
\begin{eqnarray}
    \mu_N^{ex} &\equiv& F^{ex} (N+1,V,T) - F^{ex} (N,V,T) \nonumber \\
&=& k_BT\ln{\bigg(\frac{\rho_{N+1}(r)}{\langle e^{-\beta\delta U(r)}\rangle}\frac{V}{N+1}\bigg)},
\label{eq:widinhom}
\end{eqnarray}
where $\rho_{N+1}(r)$ is the  density at the attempted insertion point of the simulation with $N+1$ particles and $\delta U(r)$ is the interaction energy of the test particle. For all practical purposes, except for very small $N$, $\rho_{N+1}(r)/(N+1)\simeq\rho_{N}(r)/(N)$, and $\mu_N^{ex}$ can be calculated with the density profile from the same simulation with $N$ particles.
We verified that, after equilibration, the excess chemical potential is indeed uniform~\cite{Frenkel2002}, even in the presence of a drop. The chemical potential $\mu$ is obtained from $\mu_N$ using the leading order finite size correction~\cite{Hong2012,Siepmann1992} $\mu\simeq\mu_N -1/2 (\partial\mu_N/\partial n)|_{n=N}$. This correction is negligibly small in most cases. In order to compute $\rho(r)$, the spherical simulation box is divided into 20 spherical shells with the same radial thickness. Then, every 100 MC step, the histogram of particles' position normalized by the shell volume enclosed in each bin is computed to get the radial density profile. The reported values of the chemical potential are averages over the 10 most external bins. 

Although the method does not require 
computing
the cluster size, for the sake of comparison, we implemented an algorithm to measure the size of the largest liquid cluster using the ten Wolde-Frenkel cluster criterion~\cite{TenWolde1998}, with the requirement of having at least 5 neighbors within a distance smaller than the Stillinger radius of $1.5 \sigma$ to be classified as a liquid-like particle. The results are compared to the critical cluster size estimated from the excess in the Appendix.

\section{Results}

Figure~\ref{fig:avgs_all} shows the values of the pressure and the excess chemical potential of a set of MC simulations for $V=2000 \sigma^3$ and 
$N$ ranging
from 1 to 600. The values of these two properties 
corresponding to
the EoS of Heier et al.~\cite{heier2018equation} are also plotted for comparison. A good agreement between the simulations and the EoS is observed for small values of $N$, corresponding to the situation where the system is dilute and the stable phase is a uniform vapor. However, the simulation values start to deviate noticeably from the EoS for $N\gtrsim 60$ when the liquid cluster is formed.

\begin{figure}[t]
     \centering
     \begin{subfigure}[b]{\linewidth}
         \centering
         \includegraphics[width=\linewidth]{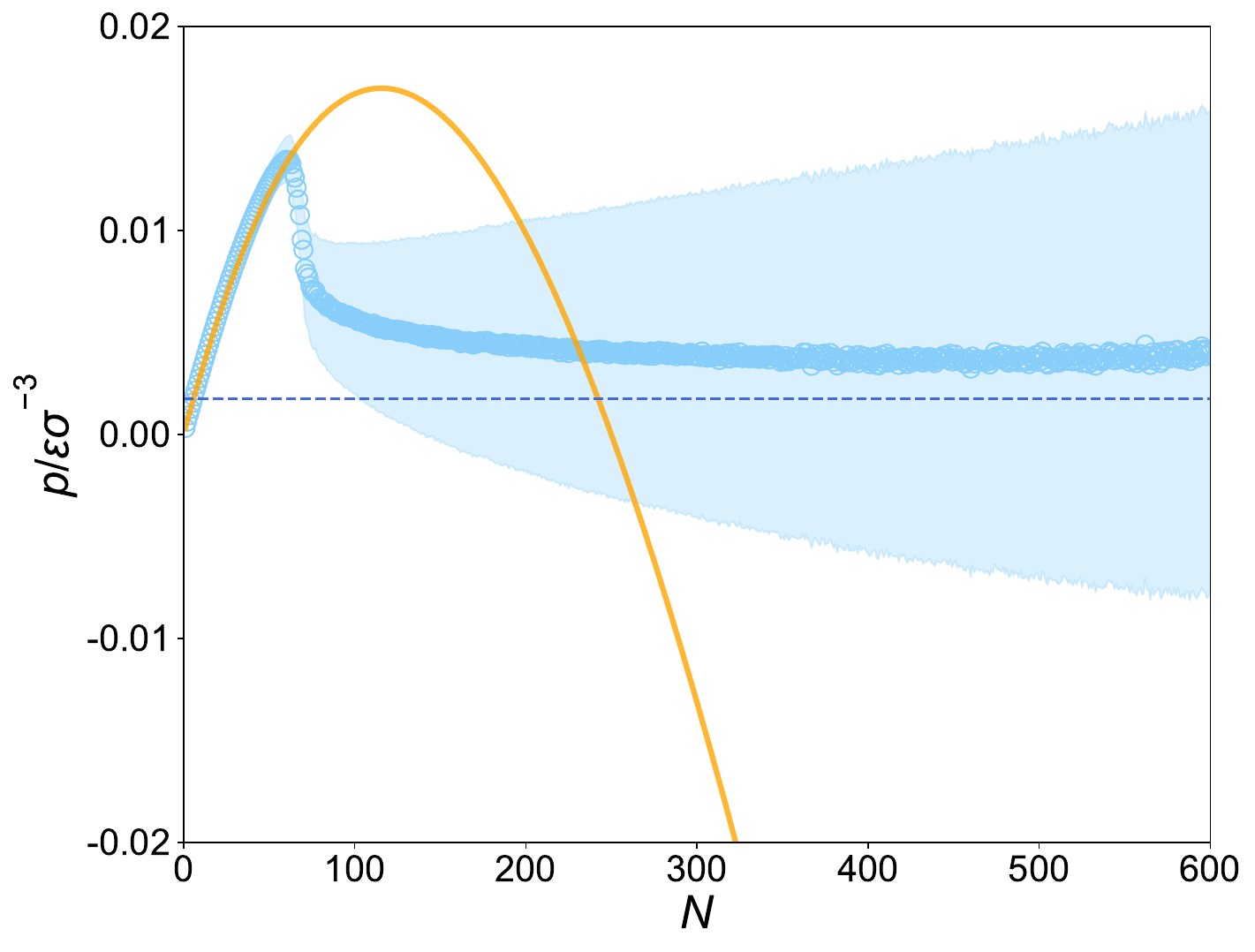}
         \caption{}
         \label{subfig:avgs_iso}
     \end{subfigure}
     \hfill
     \begin{subfigure}[b]{\linewidth}
         \centering
         \includegraphics[width=\linewidth]{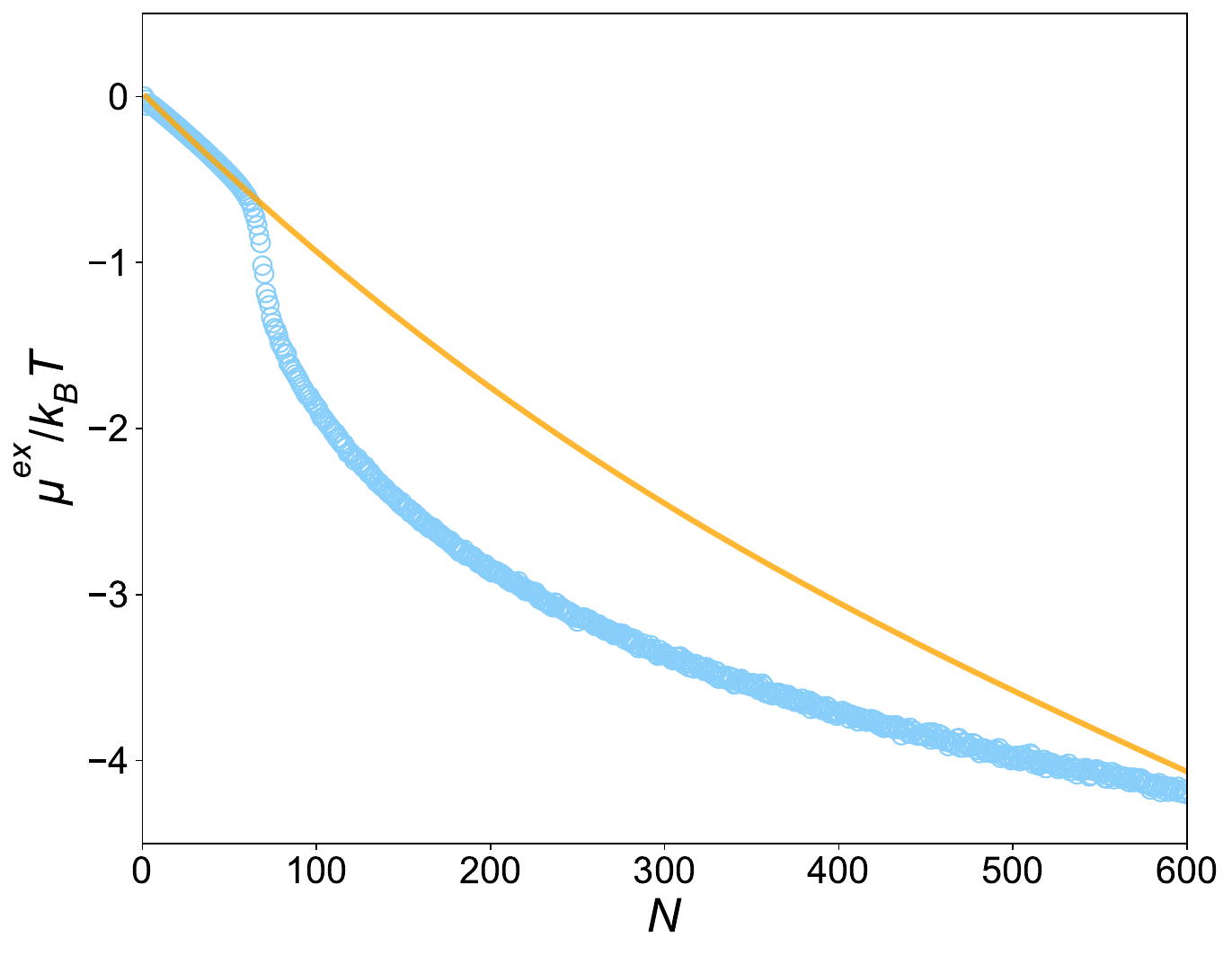}
         \caption{}
         \label{subfig:avgs_mu}
     \end{subfigure}
        \caption{\footnotesize{Average values of the (a) pressure $p$ and (b) excess chemical potential $\mu_{ex}$ 
        as a function of the number of particles $N$ in the simulation box, for $V=2000 \sigma^3$ and $T=0.625$. The dots are the results of the simulations,  and the lines represent the values of the EoS for the homogeneous fluid by Heier et al \cite{heier2018equation}. The shaded area depicts the standard deviation and the dashed blue line in (a) represents the value of the pressure at coexistence. }}
        \label{fig:avgs_all}
\end{figure}

\begin{figure}[t]
\centering
\includegraphics[width=\columnwidth]{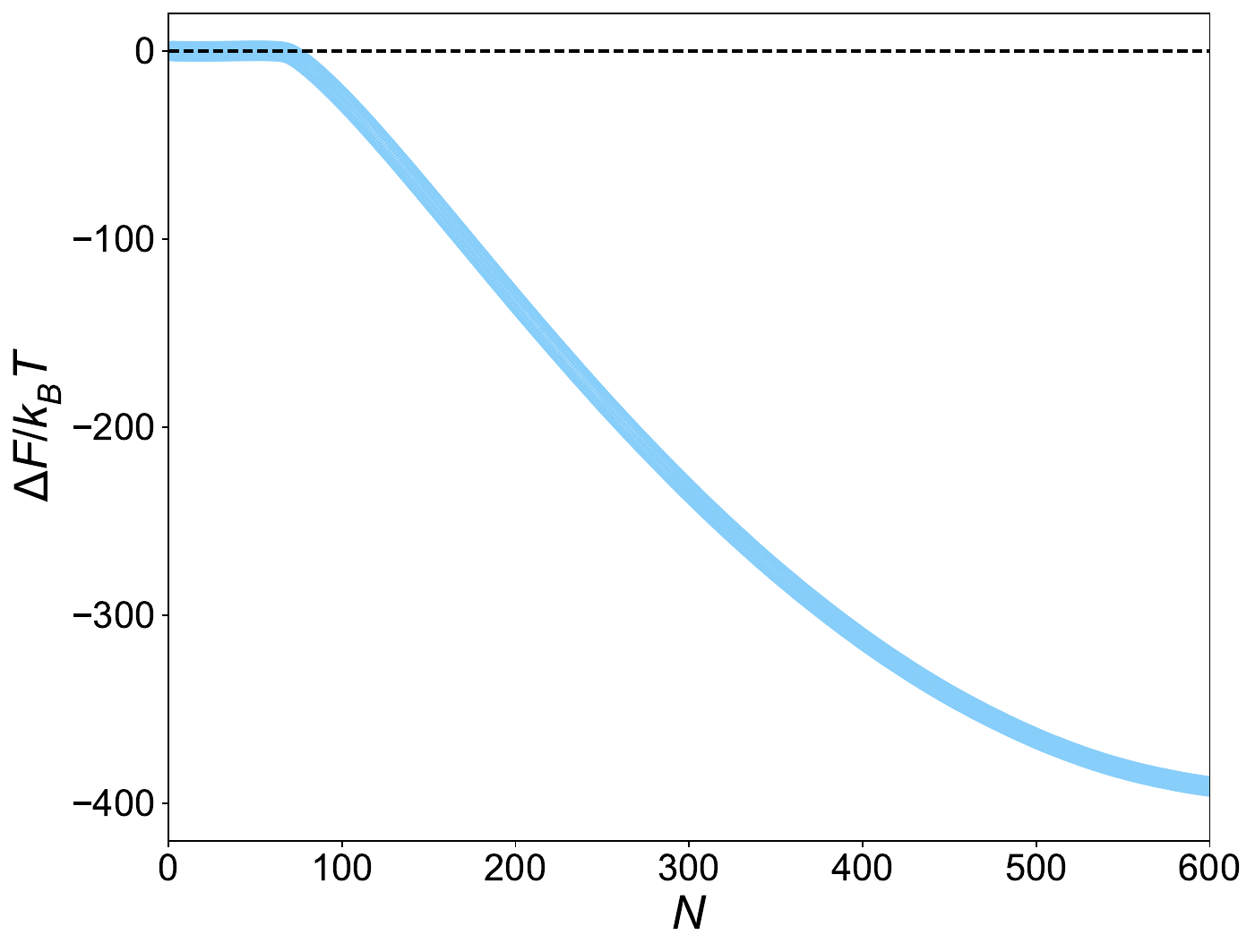}
\caption{\footnotesize{Helmholtz free energy of formation of the stable cluster as a function of the total number of particles in the simulation box $N$, for $V=2000 \sigma^3$ and $T=0.625$. The estimated error bars are smaller than the symbol size.}}
 \label{fig:DFvsN}
\end{figure}

By numerically integrating the chemical potential using Eq.~(\ref{eq:DeltaF}) with $N_{min}=1$ and a given number of molecules $N$, we obtain the Helmholtz free energy difference $\Delta F(N,V,T)$ between the simulated system and a uniform vapor. This allows us to detect the formation of a droplet in the system and to infer its stability.
Specifically, at constant temperature and volume, we have
\begin{equation}
    \Delta F(N,V,T) = \int_{N_{min}}^N \left[ \mu^{ex}(N') - \mu_0^{ex}(N')\right] dN'  , 
   \label{eq:DeltaF_constant_volume}\end{equation}
where $\mu^{ex}(N')$ is the excess chemical potential measured in the simulation with $N'$ particles, and $\mu_0^{ex}(N')$ is the excess chemical potential of the uniform vapor with density $\rho_0=N'/V$ at a temperature $T$ obtained from the EoS~\cite{heier2018equation}. 
The resulting $\Delta F$  
is shown in Fig.~\ref{fig:DFvsN} as a function of the total number of molecules $N$ in the simulation box. 
We see that $\Delta F$ is zero for $N$ small, since the system is dilute and the vapor is the stable phase,
indicating that no cluster is formed. As $N$ increases, we start to get negative values of $\Delta F$, characteristic of a stable cluster coexisting with its vapor at a chemical potential $\mu(N)$. 
This stable cluster in the canonical ensemble 
satisfies the equilibrium conditions given by Eqs.~(\ref{Eq:equal_chem}) and (\ref{Eq:Laplace}).  If instead of fixing the number of particles, we fix the value of the chemical potential in the simulation to $\mu(N)$, this cluster will correspond to the critical cluster in the $\mu V T$ ensemble, which is the only solution of Eqs.~(\ref{Eq:equal_chem}) and (\ref{Eq:Laplace}) under those constraints. 

\begin{figure}[t]
\centering
\includegraphics[width=\columnwidth]{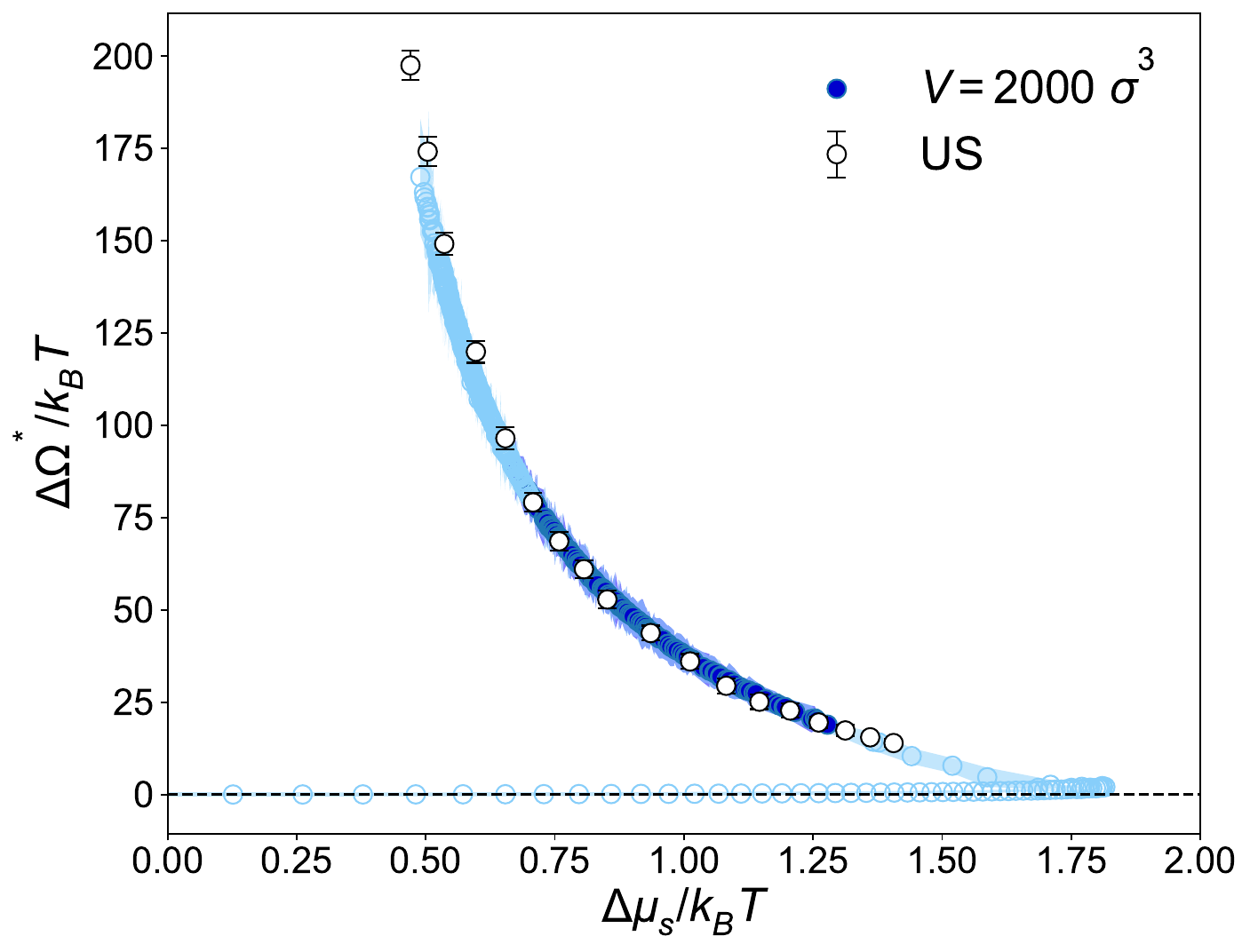}
\caption{\footnotesize{Grand free energy of formation of the critical cluster, $\Delta \Omega^*$, as a function of the supersaturation $\Delta \mu_s$, for $V=2000 \sigma^3$ and $T=0.625$. The empty blue symbols correspond to all simulated values of $N$ in the range from $N=1$ to $N=600$. The filled dark blue symbols indicate the range of values that are expected to be more accurate.  The black empty circles represent the value of the Gibbs free energy of formation as a function of the supersaturation obtained using Umbrella Sampling simulation data from Ref. \onlinecite{Aasen2023}. The shaded area represents the estimated standard deviation in $\Delta \Omega^{*}$}. }
 \label{fig:DGvsmu}
\end{figure}

According to our method, the free energy of formation of such {\it critical} cluster is obtained by means of Eq.~(\ref{eq:DeltaG_simu}). Notice that the computation of $\Delta F$ and the EoS itself are not required to obtain the height of the nucleation barrier $\Delta \Omega^*$.
The empty blue symbols in Fig.~(\ref{fig:DGvsmu}) represent the result of this 
calculation,
plotting parametrically $\Delta \Omega^*$ versus the supersaturation defined as $\Delta\mu_s(N)\equiv\mu(N)-\mu_{coex}$, where $\mu_{coex}$ is the chemical potential at coexistence given by the EoS~\cite{heier2018equation}.
For small $N$, the density is low and the system remains in the vapor phase. In this case $\Delta \Omega^*$ is essentially zero, signaling the absence of any liquid cluster. As we keep increasing $N$, a significant nucleation barrier appears at $\Delta\mu_s\approx 1.7~k_B T$, and gets larger as $N$ increases and $\Delta \mu_s(N)$ decreases. The values of $\Delta \Omega^*$  in this branch correspond to the free energy of formation of increasing large critical clusters surrounded by a vapor with a supersaturation $\Delta\mu_s$. The first few points of this barrier at the largest values of supersaturation are noisy and should not be considered, since they correspond to the evaporation transition of the cluster in the canonical ensemble\cite{Reguera2003}. Moreover, the points with the higher values of  $\Delta \Omega^*$ are also not very accurate since they correspond to 
a large cluster confined in a volume so small that  
the density profile cannot reach the value of the bulk metastable vapor density at its periphery. Therefore, as a guiding principle, accurate values of the work of cluster formation are expected in the region of $N$'s between the formation of truly stable clusters ($\Delta F<0$) and densities $N/V$ smaller than $\rho_{EoS}(p_{coex})$, i.e. the value of the density from the unstable branch of the EoS having the same pressure as the coexistence value. (For densities larger than this one could think of having a stretched fluid rather than a supersaturated vapor as a putative reference metastable state). For $V=2000 ~\sigma^3$, this range corresponds to $75<N<250$ and it is indicated in Fig.~(\ref{fig:DGvsmu}) with solid dark blue symbols.

Notice that the plot in Fig.~(\ref{fig:DGvsmu}) is not a free energy landscape for a fixed supersaturation.  
Each point corresponds to the nucleation barrier at one supersaturation. 
In this way,
with this set of simulations we 
obtain a whole set of barrier heights in a large range of supersaturations, that will require several different simulations by a standard US technique.
It is remarkable that, in this example, only up to 250 molecules are needed for getting 
reliable values of nucleation barriers between $20$ and $80~k_BT$ for a wide range of supersaturations. This feature makes 
the FRESC method
particularly attractive 
for studying complex systems or intermolecular potentials that are not easy to simulate with other techniques.

As a side benefit, since in each simulation the cluster that is formed is stable, its properties can be sampled with a lot of accuracy limited only by the statistic sampling. 
For instance, some examples of radial density profiles and the size of the stable cluster according to the ten Wolde-Frenkel cluster criteria are shown in the Appendix.

\subsection{Comparison with Umbrella Sampling}

In order to test the FRESC method, we contrasted our results with previous data from Umbrella Sampling (US) Monte Carlo simulations in the $NPT$ ensemble for the same LJTS($2.5 \ \sigma$) fluid, evaluating the Gibbs free energy of formation of the critical cluster for different values of the pressure \cite{Aasen2023}, at reduced temperature of $T=0.625$. 
Those simulations required the use of a cluster criteria and were performed with 10000 particles using 40 umbrella windows to generate each point. In order to make a comparison, the values of the vapor pressure reported in those simulations were converted to the corresponding chemical potential and supersaturation using the EoS for the homogeneous vapor \cite{heier2018equation}. 
Figure~\ref{fig:DGvsmu} shows the comparison between the two datasets. The figure reveals that the free energies of formation are in excellent agreement in the whole range of supersaturations, and are even close in the regions where the simulation results are expected to be not so accurate (corresponding to the empty blue circles). 

It is important to emphasize that the FRESC method does not rely on the validity of   Classical Nucleation Theory,  nor require the use of any cluster criteria or reaction coordinate. But of course, we can use one a specific criteria to determine the size and properties of the stable cluster, as it is shown in the Appendix.

\section{Scaling with the volume}

It is also important to analyze how the results depend on the value of the volume $V$ used in the simulations. Accordingly, we have performed simulations for different values of $V$ at the same temperature $T=0.625$.

\begin{figure}[t]
        \centering
         \includegraphics[width=\linewidth]{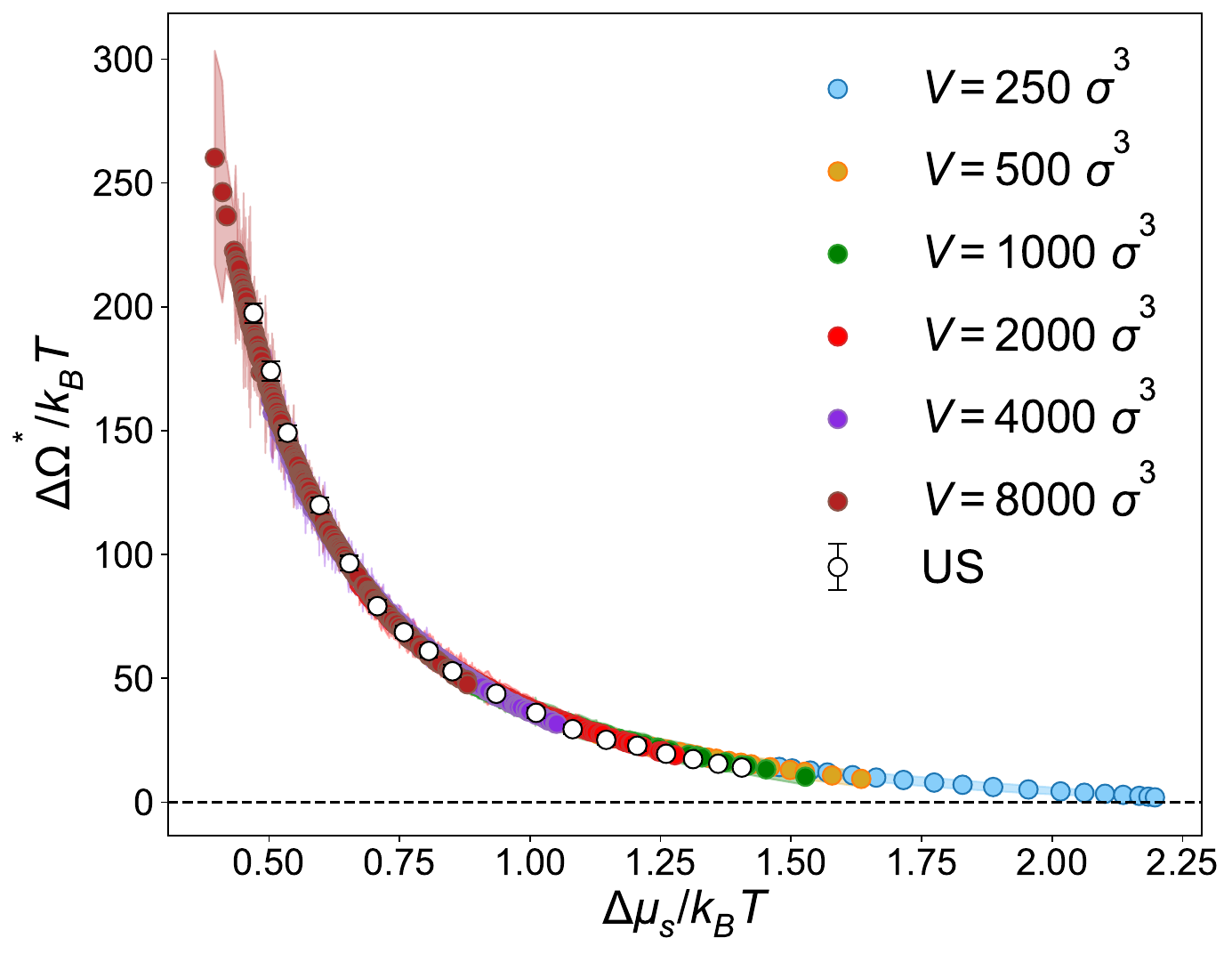}
        \caption{\footnotesize{Free energies of formation of the critical cluster, $\Delta \Omega^*$ as a function of the supersaturation $\Delta \mu_s$, at $T=0.625$ %and (a) $V=125 \sigma^3$, %(b) $V=250 \sigma^3$,  and (b) 
        for different simulations with  $V=250~\sigma^3$, $V=500~\sigma^3$, $V=1000~\sigma^3$, $V=2000~\sigma^3$, $V=4000~\sigma^3$, and $V=8000~\sigma^3$. The filled circles are the results of the simulations, and the open circles represent the values of previous US simulations of Ref.~\onlinecite{Aasen2023}. The shaded area represents the estimated standard deviation in $\Delta \Omega^{*}$. }}
        \label{fig:V_scaling}
\end{figure}

Figure~\ref{fig:V_scaling} shows the results for the free energy of formation of the critical cluster $\Delta \Omega^*$ for different volumes in the range from $V=250~\sigma^3$ to $V=8000~\sigma^3$. In all cases, only values of the nucleation barrier in a range of sizes $N$ are shown, since for small $N$
no cluster forms and for large $N$ the results might not be accurate since there are very few vapor molecules.
More specifically, Fig.~\ref{fig:V_scaling} shows the results obtained using $V=250 \sigma^3$, with $N$ from 14 to 30; $V=500 \sigma^3$, with $N$ from 35 to 60; $V=1000 \sigma^3$, with $N$ from 50 to 120; $V=2000 \sigma^3$, with $N$ from 75 to 250; $V=4000 \sigma^3$, with $N$ from 120 to 500; and $V=8000 \sigma^3$, with $N$ from 200 to 900. Remarkably, despite the 
wide range of volumes and particle numbers, all simulations nicely collapse onto
a single curve, that coincides with the results obtained from US simulations at a much larger computational cost (40 windows with $10^4$ molecules for each point in the US simulations vs. a set of typically 200 simulations with less than 500 molecules in the 
FRESC
method without any biasing, cluster criteria or complex reweighting reconstruction involved).
Notably, these simulations also yield an estimate of the nucleation barriers at extremely high supersaturations very close to the spinodal limit (which, according to the EoS is predicted to be at $\Delta \mu_s= 2.0~k_B T$), in a region where it was not possible to get reliable results using US simulations.

It is specially important to highlight that for the barrier heights of interest for condensation at experimental conditions, the typical critical cluster sizes are on the order of 100 molecules.  
These clusters could be characterized
very efficiently with the new method using simulations with up to 200 or 300 molecules.

\section{Conclusions}

We have presented a new simulation method to evaluate the free energy of formation of critical clusters, which is the crucial ingredient in nucleation theory. The FRESC method is simple, efficient, uses a small number of molecules $N$, is not based on the validity of CNT, and does not require any reaction coordinate or order parameter. The method relies on the use of the $NVT$ ensemble to generate a stable cluster that would correspond to the critical cluster in the $NPT$ or $\mu VT$ ensemble of interest for nucleation.

The method has been illustrated by performing Monte Carlo simulations for a system of  particles interacting with a Lennard-Jones potential truncated and shifted at $2.5 \ \sigma$, inside a spherical volume at reduced temperature of $T=0.625$. We have shown that the results of the new method compare well with MC Umbrella Sampling simulations, at a much reduced complexity and computational cost. We can even explore extremely high supersaturations close to the spinodal where US or other methods cannot be easily used.

The accuracy of the method can be further improved by 
going beyond the approximation $\mathcal{E}_0\equiv  F_0^{ex} +p_0^{ex}V -\mu_0^{ex} N =0$, used in Eq.~(\ref{eq:DeltaG}), which is strictly valid only in the 
thermodynamic limit. Indeed, in the simulations we find that  $\mathcal{E}_0$ is different from zero even in the uniform vapor phase. This correction is small in the present case, but can be higher at temperatures close to the critical point. A simple and approximate way to introduce this correction would be to follow the same strategy performed in Ref. \cite{Neimark2005c} and remove from Eq.~(\ref{eq:DeltaG}) the value of $\mathcal{E}_0$ from the metastable vapor branch (the lower branch with empty symbols in Fig.~\ref{fig:DGvsmu}) having the same supersaturation. Another computational alternative would be the use of the restricted ensemble\cite{Corti1994} to stabilize the metastable vapor phase for large densities and estimate $\mathcal{E}_0$. However, those are just approximations and a careful estimation of the finite size correction for very small sizes following the ideas of Hill's small systems thermodynamics \cite{Hill,Brten2021a,Brten2021b} will be performed in the future. The FRESC method in its present form is also not suited to simulate very large systems or cluster sizes. In that case, the transition between the vapor and the droplet in the canonical ensemble will show a hysteresis loop that should be corrected in order to evaluate the work of formation from thermodynamic integration. Moreover, small clusters cannot be stabilized in very large volumes\cite{Reguera2003,Wilhelmsen2014}.

Although we have applied the method to study condensation in a simple Lennard-Jones fluid, it can be applied to much more complicated substances and complex intermolecular potentials inaccessible for other simulation techniques that would require a much larger number of particles.

The method can also be straightforwardly extended to other nucleation phenomena like cavitation, bubble formation, crystallization or binary or heterogeneous nucleation.
The FRESC technique will open the door to accurate simulations and predictions of nucleation barriers and rates for substances of industrial or atmospheric interest.

\begin{acknowledgments}
This paper is dedicated to Carlos Vega on the occasion of his 60th birthday. His passion for science and his human values are an inspiration for all of us. 
We wish to acknowledge fruitful discussions with V. Molinero and P. Montero de Hijes, and funding from grants PID2021-126570NB-I00 and PID2024-156516NB-I00 financed by MICIU/AEI/10.13039/501100011033 and FEDER/UE.
\end{acknowledgments}

\noindent {\bf{Conflicts of Interest}}: The authors declare no conflict of interest.

\section*{Data Availability Statement}

The data that support the findings of this study are available from the corresponding author upon reasonable request.

\appendix*
\section{Critical cluster size and density profiles}

\begin{figure}[t]
\centering
\includegraphics[width=\columnwidth]{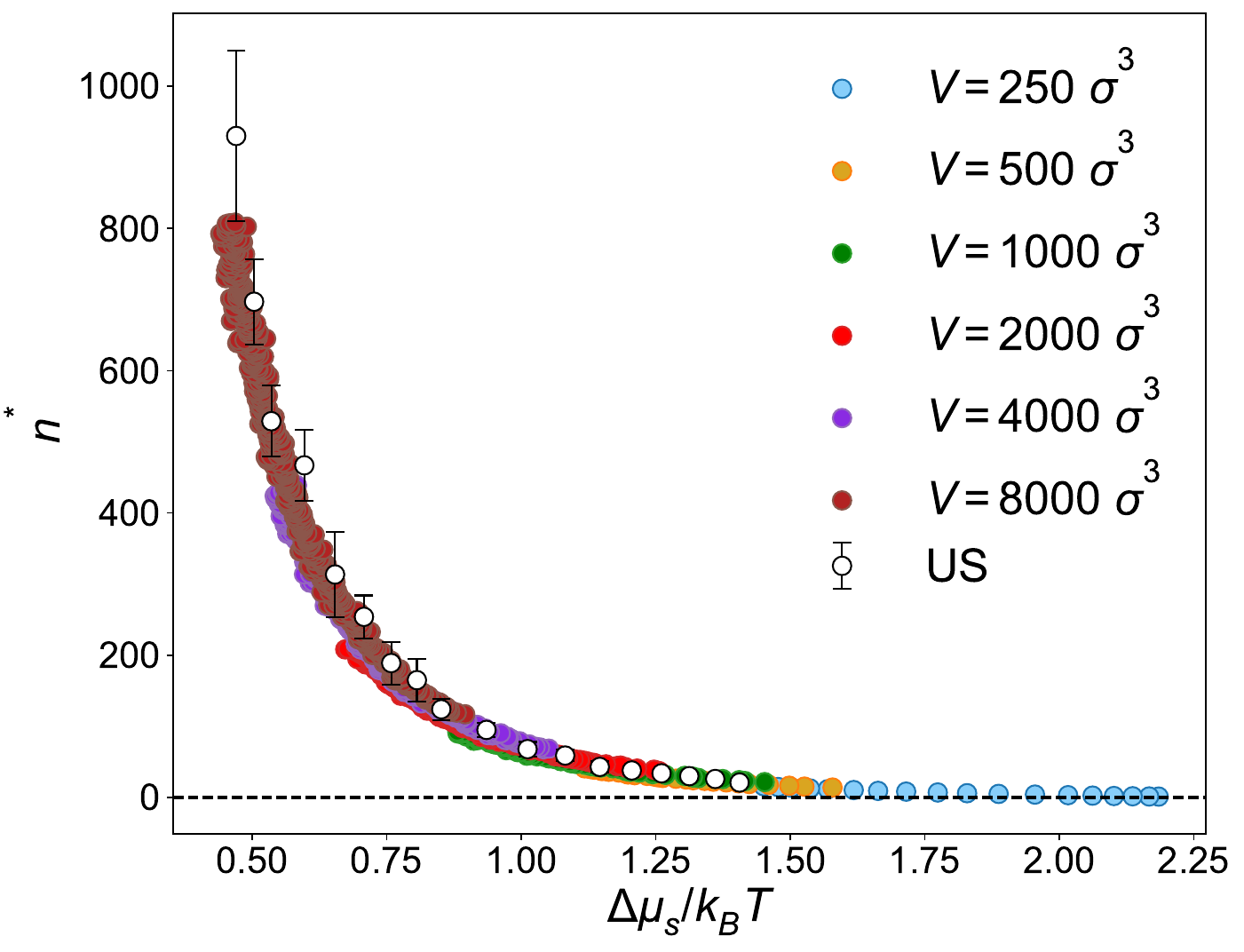}
\caption{\footnotesize{ Comparison between the critical cluster size for US data in Ref. \cite{Aasen2023} and our simulations, versus the supersaturation, for a LJTS($2.5 \ \sigma$) fluid at $T=0.625$.} }
\label{fig:ndemu}
\end{figure}

Figure~\ref{fig:ndemu} shows the size of the critical cluster versus the supersaturation, determined in our simulations using the ten Wolde-Frenkel cluster criteria with a Stillinger radius 1.5 $\sigma$ and requiring at least 5 neighbors for a particle to be identified as liquid-like. We compare the results with those of the US simulations of Ref.~\onlinecite{Aasen2023}, obtaining a good agreement.

\begin{figure}[t]
\centering
\includegraphics[width=\columnwidth]{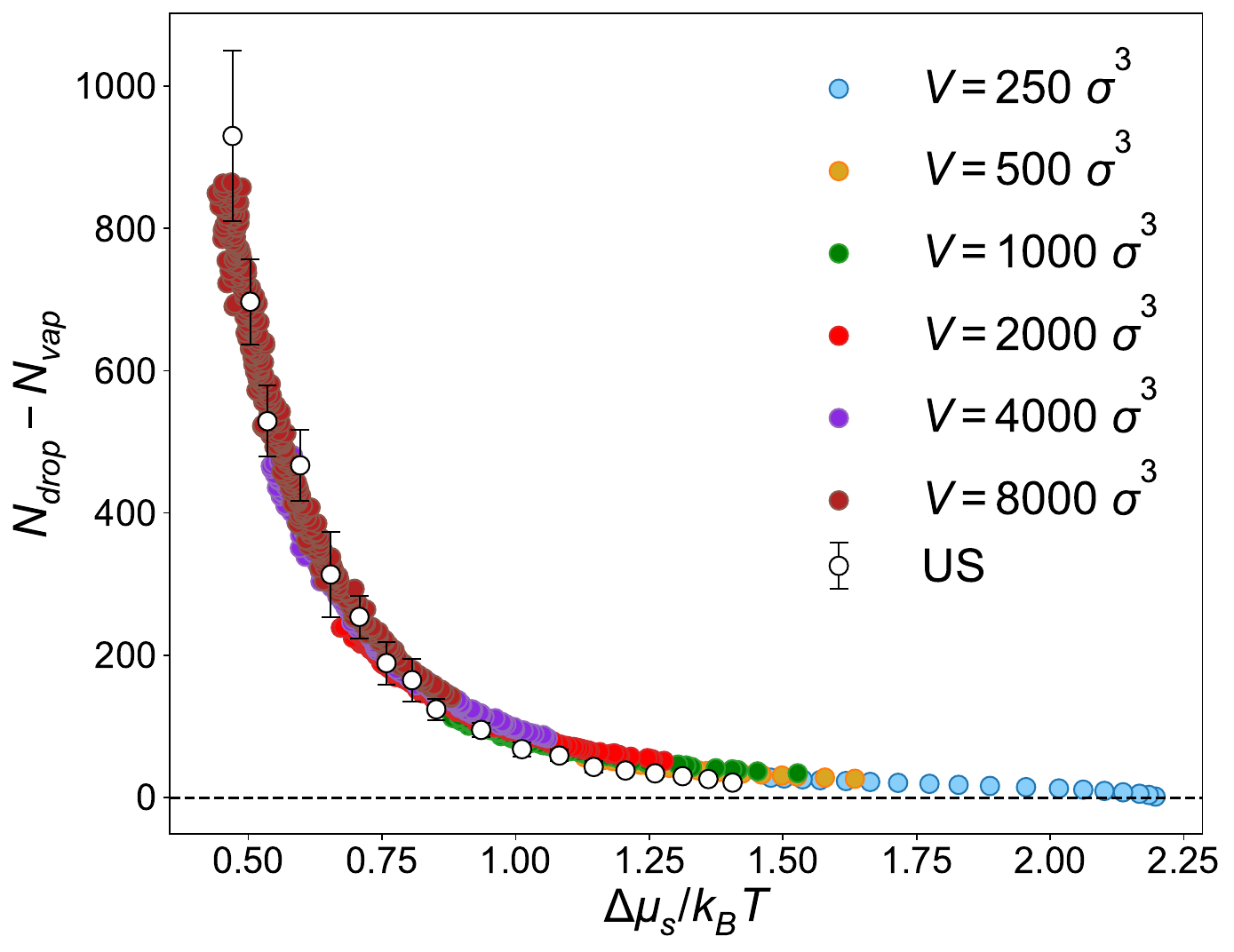}
\caption{\footnotesize{Comparison between the excess number of molecules $N_{drop}-N_{vap}$ obtained as indicated in Fig.~\ref{fig:supersaturation} and the critical cluster size for US data in Ref. \cite{Aasen2023}, versus the supersaturation, for a LJTS($2.5 \ \sigma$) fluid at $T=0.625$.}  }
\label{fig:excess}
\end{figure}

\begin{figure}[t]
\centering
\includegraphics[width=\columnwidth]{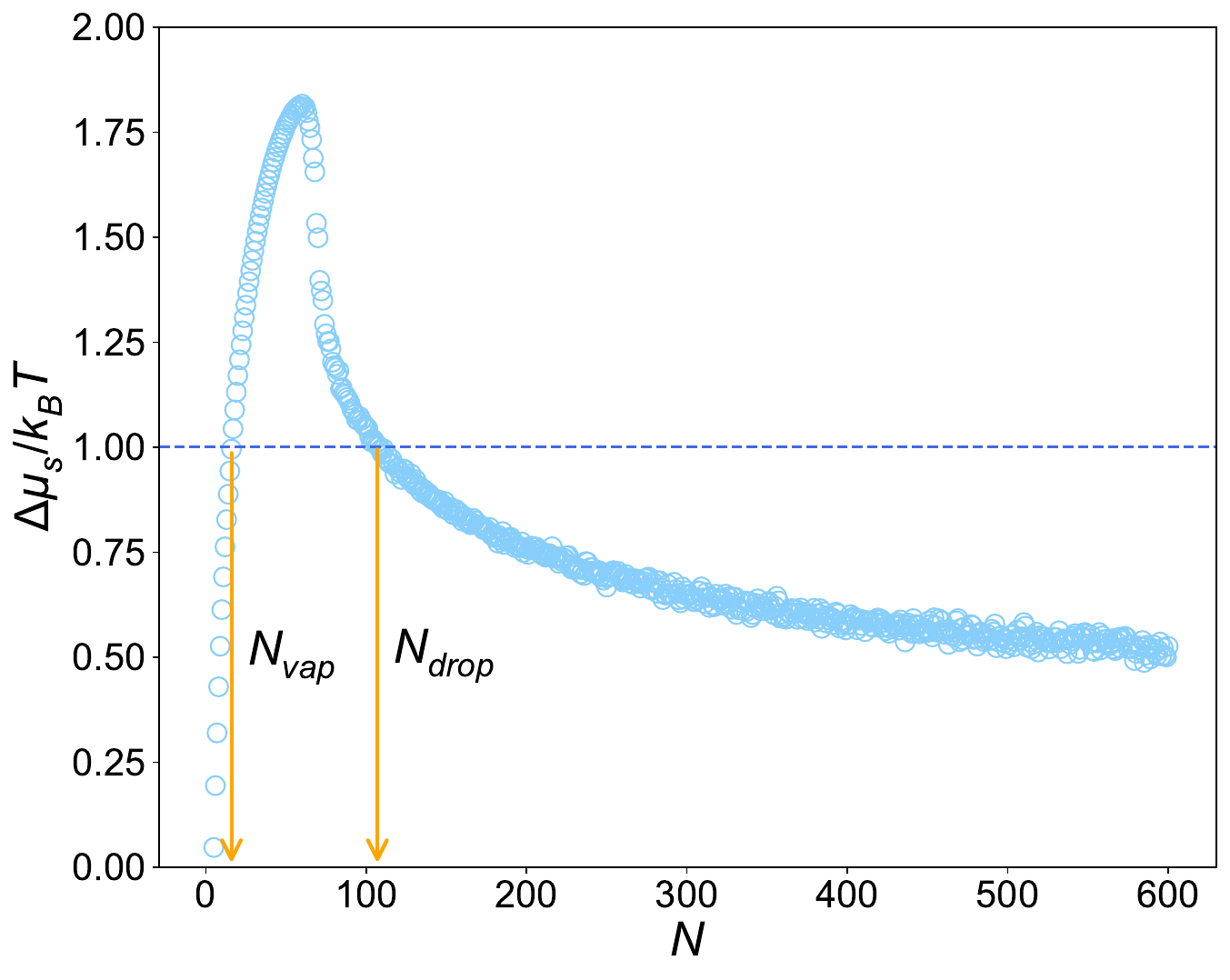}
\caption{\footnotesize{Value of the supersaturation $\Delta \mu_s$ obtained from the simulations performed with $V=2000\sigma^3$ and different number of molecules $N$. For any specific value  of the supersaturation $\Delta \mu_s$, indicated by the blued dashed line, there are two simulations sizes $N_{drop}$ and $N_{vap}$, corresponding to a drop surrounded by vapor, and uniform vapor, respectively, both having the same supersaturation. The difference  $N_{drop}-N_{vap}$ is a measure of the molecular excess of the drop plotted in Fig.~\ref{fig:excess}.}}
\label{fig:supersaturation}
\end{figure}

The size of the critical cluster can also be estimated without resorting to any cluster criterion by evaluating the excess number of molecules between a simulation with a drop $N_{drop}(\Delta \mu_s)$ and the simulation with a uniform vapor $N_{vap}(\Delta \mu_s)$ having the same chemical potential $\Delta \mu_s$ (see Fig.~\ref{fig:supersaturation}). The resulting values are represented in Fig.~\ref{fig:excess}, compared against those of the US simulations. We can see that for the smaller clusters, the molecular excess is consistently larger than the sizes identified by the ten Wolde-Frenkel cluster criteria.

\begin{figure}[t]
\centering
\includegraphics[width=\columnwidth]{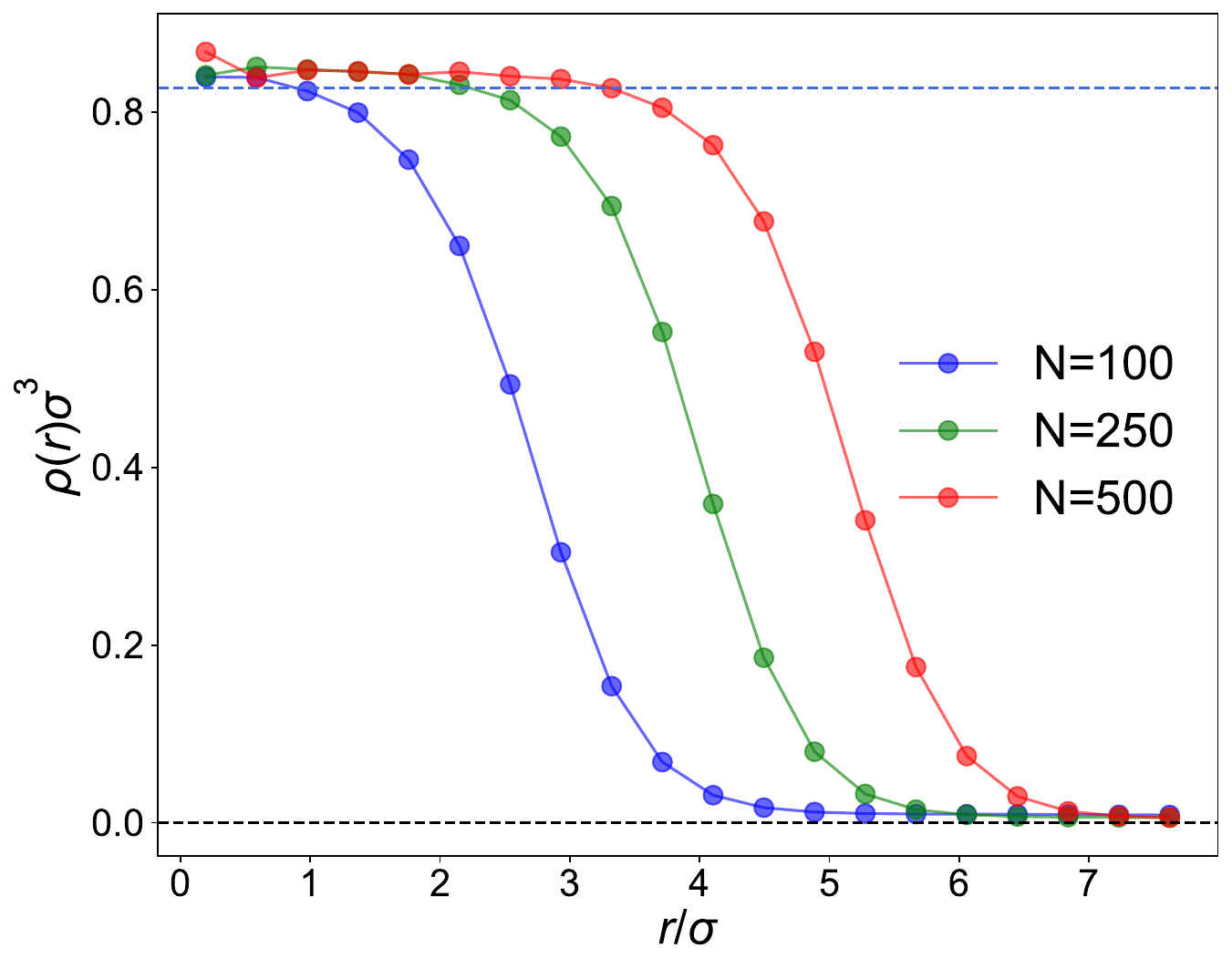}
\caption{\footnotesize{Radial density profiles for a volume $V=2000\ \sigma^3$, and sizes $N=100$, $N=250$ and $N=500$ at $T=0.625$. The bulk liquid density of the LJTS($2.5 \ \sigma$) at $T=0.625$ is shown as a reference.  }}
\label{fig:density_profiles}
\end{figure}

Finally, Fig.~\ref{fig:density_profiles} illustrates the radial density profile of three stable clusters for $V=2000~\sigma^3$ and $N=100$, $N=250$ and $N=500$. We can see that for $N=500$, the volume of the container is similar to the droplet size and the density profile cannot reach that of the bulk metastable vapor at the periphery.  
Therefore, if one is also interested  in the microscopic properties of the critical cluster, 
the fact that the canonical ensemble stabilizes the cluster enables accurate sampling of those properties.

\nocite{*}
\bibliography{references}% Produces the bibliography via BibTeX.

\end{document}